\documentclass[runningheads]{llncs}
\usepackage[utf8]{inputenc}

\usepackage{multirow}
\usepackage{array}
\usepackage[T1]{fontenc}
\usepackage{amsmath}
\usepackage{amsfonts}
\usepackage{bm} 
\usepackage{booktabs}
\usepackage{pifont}

\usepackage[compatibility=false]{caption}
\usepackage{subcaption}

\usepackage{graphicx}

\usepackage{verbatim}
\newcolumntype{x}[1]{>{\centering\arraybackslash}p{#1}}

\begin{document}
\title{cSVR: Convolutional Slice-to-Volume Reconstruction}
%
\author{Margherita Firenze\inst{1}\textsuperscript{(\ding{41})}\and
Sean I. Young\inst{2}\and
Clinton Wang\inst{1}\and
Elfar Adalsteinsson\inst{3}\and
HyukJin Yun\inst{4,5}\and
Patricia E. Grant\inst{6}\and
Kiho Im\inst{6}\and
Polina Golland\inst{1}}
\authorrunning{M. Firenze et al.}
\institute{Computer Science and Artificial Intelligence Lab, MIT, Cambridge, USA \email{mfirenze@csail.mit.edu} \and
Athinoula A. Martinos Center for Biomedical Imaging, Massachusetts General Hospital and Harvard Medical School, Charlestown, USA \and
Research Laboratory of Electronics, MIT, Cambridge, USA \and
Department of Radiology, Children's Mercy Hospital, Kansas City, USA \and
School of Medicine, University of Missouri-Kansas City, Kansas City, USA \and
Boston Children's Hospital and Harvard Medical School, Boston, USA}
\maketitle              
\begin{abstract}
Unpredictable fetal motion during MRI scans can result in oblique 2D slices that are difficult to interpret and often leads to prolonged scan times. Slice-to-volume reconstruction (SVR) methods align 2D oblique slices from multiple slice stacks into a coherent 3D volume. We propose a convolutional network that predicts slice poses by performing multiscale, unrolled optimization and further refines them and super-resolves the volume using model-based optimization. Our convolutional feed-forward SVR network achieves accuracy on par with state-of-the-art results while offering significant speedup, paving the way for scanner-side use of SVR. The code is available publicly:\\
\url{https://github.com/MedicalVisionGroup/cSVR}.


\end{abstract}

\section{Introduction}
Fetal magnetic resonance imaging (MRI) enables clinical diagnosis and scientific study of the developing brain~\cite{mri_fetal_medicene,mri_brain_abnorm,fetal_mri} but is complicated by fetal motion. Fast MRI sequences limit motion artifacts in the acquired 2D images (slices)~\cite{fetal_mri_sequence} but significant motion during the  ``cool off’’ period between slice acquisitions leads to uneven coverage and oblique slices. To support clinical assessment, image stacks are reacquired until all slices are minimally oblique. This results in long, uncomfortable scan times for the mother and a high cognitive burden for radiologists who evaluate the images.

These problems can be overcome by slice-to-volume reconstruction (SVR) methods that co-align all slices in 3D and super-resolve a coherent volume consistent with the acquired data~\cite{Gholipour2010,svrtk,Rousseau2006,nesvor1}. SVR is widely used in research~\cite{CRL,facial_modeling}, but clinical adoption remains limited: existing pipelines integrate SVR into clinical workflows~\cite{Uus2025-fa}, but state-of-the-art runtimes of several minutes disrupt real-time review by radiologists. In addition to enabling clinical use, fast SVR also has the potential to reduce total imaging time and improve image quality by guiding decision-making during acquisitions: when to stop acquiring new data because the brain coverage is complete, and which orientation to prescribe for the next stack.

Early methods framed SVR as an optimization problem and solved it by alternating between slice pose estimation and volume reconstruction~\cite{ebner,ebner2020automated,Gholipour2010,Jiang2007,kainz_2015,svrtk,Rousseau2006,tourbier2015efficient}. Like many other gradient descent solutions, this approach is sensitive to poor initialization and struggles when large motion is present. Recent deep learning methods directly predict slice poses using CNNs to improve speed and robustness~\cite{hou2018,pei,mohseni2018}. Adding more task-specific priors, such as slice and stack encoding or time modeling improves the accuracy of registration~\cite{svr_mamba,svort}. Unfortunately, the slice poses estimated by existing feedforward methods are not accurate enough and are thus usually used to initialize slower, optimization-based solutions. Fully neural approaches optimize implicit neural representations at inference time to jointly model volume and poses, achieving state-of-the-art reconstructions, but require long runtimes and specialized GPU hardware, even at inference~\cite{meta_svr,sufficient2025,nesvor1}.

Our work draws on a recently proposed fully convolutional single-stack SVR method, which employs a U-Net to alternate between free-form slice registration and volume reconstruction, followed by a U-Net for super-resolution and inpainting~\cite{sean}. Performing SVR on a single stack is uncommon, as multiple stacks are acquired in clinical practice and one stack does not adequately cover all areas of the brain, resulting in reconstructions with holes. 

We build on this convolutional approach to handle multiple stacks acquired in real-world clinical settings by explicitly modeling stack orientation that is automatically inferred from the data. Rather than use inpainting networks that might introduce or remove important anatomical details, we integrate our network with model-based reconstruction to refine poses and perform super-resolution. We model the slice pose as a free-form displacement field that is later projected on to the space of rigid poses. In this paper, we demonstrate high quality rigid SVR in less than 20s with $\sim$1s slice registration, offering significant speedup over state-of-the-art fast SVR solutions~\cite{svort}.

\section{Methods}

Given a collection of slices $\bigl\{ I_n \bigr\}$, each defined on a discrete 2D pixel grid~$\Omega$, SVR estimates the slice poses $\bigl\{ F_n \bigr\}$ and the underlying coherent volume~$V$~\cite{Gholipour2010,svrtk,Rousseau2006}. Classical methods minimize self-consistency loss based on the differences between the predicted slices and the acquired data $\bigl\{ I_n \bigr\}$:

\begin{equation}
\label{eq:classical_eq}
(\hat{V}, \{ \hat{F}_n \})
=
\operatorname*{argmin}_{V, \{ F_n \}}
\sum_{n}
\left\| \mathcal{V}(F_n, V) - I_n \right\|^2.
\end{equation}
\noindent
where $\mathcal{V}(F,V)$ is the MRI forward model that generates the predicted slice based on pose $F$ and volume $V$. This optimization is typically solved via coordinate descent by alternating between the estimation of $V$ (reconstruction) and $\bigl\{ F_n \bigr\}$ (registration). Slices are typically masked as a pre-processing step to only include the intracranial content.  A reference stack is usually chosen and all other stacks are registered to the reference stack. The slice poses are then initialized using their respective stack pose. In addition to long run time, this solution results in reconstructions that are not guaranteed to be aligned with the standard radiological planes and require further post-processing for proper alignment.

\begin{figure}[t!]
   \centering
   \includegraphics[width=1.0\textwidth]{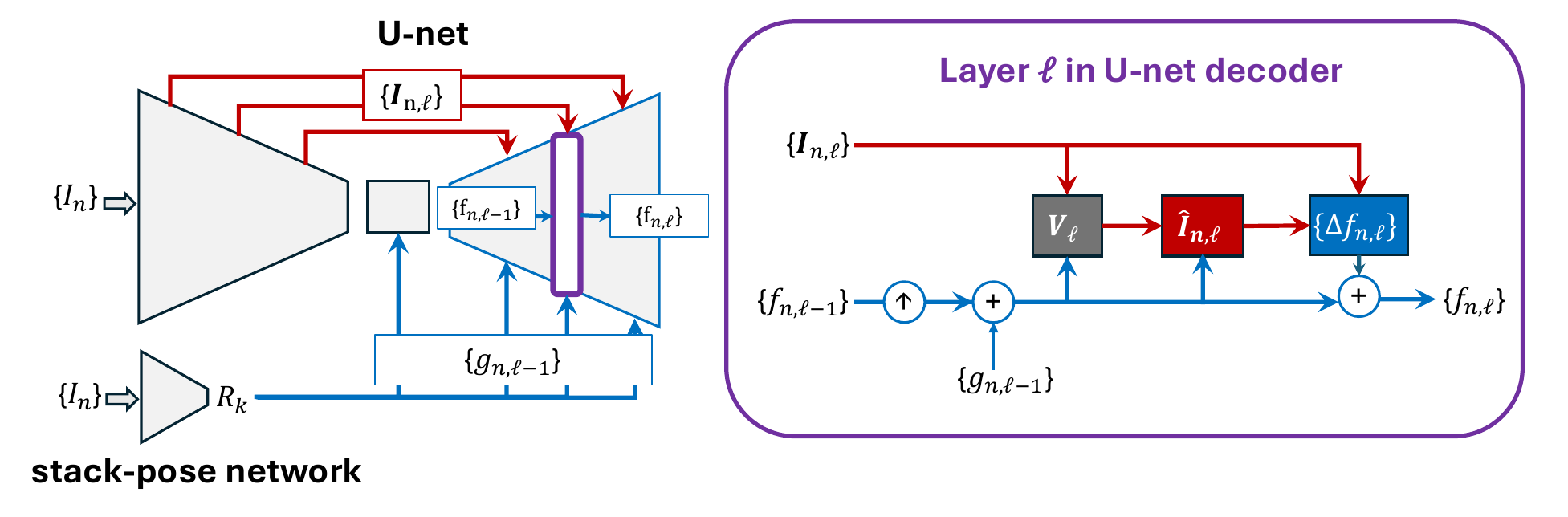}
     \caption{\textbf{cSVR pose estimation:} Our method estimates slice poses using two networks: the stack-pose network estimates the bulk pose $g_{n,\ell}$ and the U-Net refines the residual field $f_{n,\ell}$. The stack-pose network estimates the orientation $R_k$ for each stack, which is used to calculate the bulk-pose estimate $g_{n,\ell}$ at layer $\ell$. The U-Net decoder alternates reconstruction (estimating feature volume $\mathbf{V}$) and slice pose estimation to refine $f_{n,\ell}$. }

   \label{fig:overview}
\end{figure}

\subsection{cSVR: convolutional SVR}

We develop a convolutional network that is trained to solve the optimization problem in Eq.~(\ref{eq:classical_eq}) as an unrolled, multiscale alignment of all slices to a volumetric pose that reflects the standard radiological orientation. We model the slice pose as a free-form transformation to support local convolution-based updates, inspired by the optical flow solutions~\cite{Horn}. We model the slice pose as a vector field $F:\Omega \rightarrow \mathbb{R}^3$, where $g: \Omega \rightarrow \mathbb{R}^3$ is a bulk pose field and $f: \Omega \rightarrow \mathbb{R}^3$ is a residual field, such that pixel $p \in \Omega$ is mapped to 3D location $F(p) = g(p) + f(p)$. We predict the bulk pose $g$ using a stack-pose network and estimate $f$ using a specialized U-Net. We initialize the residual field $f$ to zero and refine it throughout the decoder layers of the U-Net by performing reconstruction and slice prediction steps similar to classical methods, but operating on (multi-channel) \textit{feature} slices $\bm{I}$ and \textit{features} volumes $\bm{V}$. In the remainder of this section, $g_{n,\ell}$ refers to the bulk pose field for slice $n$ sub-sampled and rescaled for layer $\ell$ of the U-Net decoder, and $f_{n,\ell}$ refers to the residual field estimate for slice $n$ computed by layer $\ell$ of the U-Net decoder (Fig.~\ref{fig:overview}).

We project the final slice pose field $F_n$ onto the space of rigid transformations~\cite{project} and use it to initialize a model-based optimization that iteratively refines both the 3D volume $V$ and rigid slice poses to optimize consistency with the acquired slices in Eq.~(\ref{eq:classical_eq})~\cite{svrtk}. We employ a GPU-accelerated implementation of this model-based reconstruction~\cite{svort}.

\noindent

\noindent
\textbf{Bulk Pose \boldmath$g_{n,\ell}$}. The stack-pose network classifies the orientation of each stack $k$ as one of six standard orientations: axial, coronal, sagittal, combined with the direction (e.g., top-to-bottom or bottom-to-top) and constructs the corresponding rotation matrix $R_k$. For each layer $\ell$ of the U-Net decoder, we construct the bulk pose field $g_{n,\ell}$ such that for any pixel $p =[p_x, p_y]^\text{T} \in \Omega$,

\begin{equation}
\label{slice param}
g_{n,\ell}(p) =  C_\ell R_k C_{\ell}^{-1} D_\ell T_n \, \begin{bmatrix} p_x \\ p_y \\ 1 \end{bmatrix},
\end{equation}
\noindent
where $T_{n}$ captures the  position (translation) of slice $n$ in its stack, $C_\ell$ is the translation that places the volume center at the origin for layer $\ell$, and diagonal matrix $D_\ell$ rescales slice pixel dimensions to volume voxel dimensions.

\noindent

\noindent
\textbf{Residual field \boldmath$f_{n,\ell}$.} Layer $\ell$ of the decoder first constructs volume $V_{\ell}$ as follows:
\begin{equation}
\label{recon_model_sean}
\bm{V}_{\ell}(x)
=
\frac{
\sum_{n}
\mathcal{V^*}\!\left(
g_{n,\ell} +  f_{n,\ell-1}^{\uparrow},\, \bm{I}_{n,\ell}
\right)(x)
}{
\sum_{n}
\mathcal{V^*}\!\left(
g_{n,\ell}  +  f_{n,\ell-1}^{\uparrow},\, \mathbf{1}
\right)(x)
},
\end{equation}

\noindent
where $\uparrow$ represents upsampling of the residual field $f_{n,\ell-1}$ to match the resolution of $g_{n,\ell}$, and $\mathcal{V^*}(F,\bm{I})$ is an operator that computes feature volumes by orienting the multi-channel features slice $\bm{I}$ according to pose field $F$ while using trilinear interpolation to handle sub-grid locations. Since multiple slices may contribute to a single 3D voxel, we normalize each voxel intensity by the sum of the corresponding interpolation weights. These weights are obtained by applying the same volume-pushing operation to a slice of ones.

Next, we compute a prediction for feature slices using the current slice pose estimates:
\begin{equation}
\label{forward model sean}
\hat{\bm{I}}_{n,\ell} = \mathcal{V}(g_{n,\ell} + f_{n,\ell-1}^{\uparrow}, \bm{V}_{\ell}),
\end{equation}
where $\mathcal{V}(F,\bm{I})$ constructs slice features $\hat{I}_n$ similar to Eq.~(\ref{eq:classical_eq}). 

Finally, we employ a learned convolutional operator $\Delta f$ that refines the residual field by comparing the predicted and acquired slices:

\begin{equation}
\label{flow_update_sean}
f_{n,\ell} =  f_{n,\ell-1}^{\uparrow} + \Delta f_\ell(\hat{\bm{I}}_n, \bm{I}_n).
\end{equation}

\noindent We note that $\mathcal{V}$ and $\mathcal{V^*}$ are based on the forward model of MRI slices; the operator $\Delta f$ is the only component of the U-Net decoder that is learned. 

\subsection{Training}

We train the stack-pose network using a cross-entropy classification loss over the six possible orientations. To train the U-Net we employ a multiscale $\mathcal{L}_2$ loss on the difference between the estimated residual field for each layer upsampled to the target resolution and the ground truth residual field $f^\texttt{GT}$:

\begin{equation}
\label{loss}
\mathcal{L}(f,f^\texttt{GT}) =  \sum_{n} \sum_{\ell} \left\|   f_{n,\ell}^{\uparrow} -f^\texttt{GT}_{n,\ell}\right\|_2^2.
\end{equation}

\noindent
We optimize over the residual field $f$ rather than the full transformation $g + f$, as parameterizing the problem with smaller-magnitude updates leads to improved optimization stability.

\subsection{Implementation Details} 

We train both models for 250,000 steps on an NVIDIA H200 GPU using Adam with an initial learning rate of $10^{-4}$ and polynomial decay~\cite{adam}. The final checkpoint is used in all experiments. The U-Net's convolutional encoder contains 5 2D layers with image dimensions $[128^2,\allowbreak 64^2,\allowbreak 32^2,\allowbreak 16^2,\allowbreak 8^2]$ and channel dimensions $[64, 128, 256, 512, 1024]$, respectively. We experimented with different layer numbers for the decoder and found that adding an additional bottleneck layer improves accuracy. We fix the decoder to contain 6 layers with volume dimensions $[8^3,8^3,16^3,32^3,64^3,128^3]$ and use this architecture in all evaluations in the next section.  We build the stack-pose estimation network using the same architecture as the first two layers of the 2D encoder of the U-Net followed by an MLP with hidden layers of sizes $[256, 128]$ and a length-6 output layer.

\section{Experiments and Results}
\noindent
\textbf{Data.} We train our models using data from two publicly available datasets, FeTA 2.1~\cite{feta} and the CRL atlas~\cite{CRL}. FeTA consists of high-quality T2-weighted SVR reconstructions (using classical methods) for 120 subjects (gestational age (GA) $20-35$~weeks, voxel size $0.8~\text{mm}^3$), and CRL includes 18 temporal volumes (GA $21-38$~weeks, voxel size $0.8~\text{mm}^3$). We train the network on 108 FeTA subjects and 18 CRL atlases. We sample random motion parameters during training to simulate imperfect plane selection and varying levels of motion. We sample rotations uniformly in the range $[-12^\circ, 12^\circ]$ and rotate the volume to initialize imperfect plane selection. We then sample 1--5 motion parameters per stack and smoothly interpolate between them using cubic B-splines, capturing both gradual and abrupt motion. Rotations follow a zero-mean normal distribution with $20^\circ$ standard deviation, and translations are uniformly sampled in $[-6.1, 6.1]$\,mm range. Rician noise and signal void augmentations are added for robustness~\cite{svort}. To train the stack-pose network, we additionally apply random $\pm 180^\circ$ rotations to simulate variability in fetal head orientation (e.g., breech vs. cephalic presentation, left- vs. right-facing, and inward- vs. outward-facing).

We evaluate our method on the same 12 held-out FeTA subjects as SVoRT~\cite{svort} and a private dataset of $18$ subjects from Boston Children's Hospital (GA $19-35$ weeks, pixel size $1-1.4$mm, slice thickness $2-3$mm). To enable realistic evaluation on synthetic examples, we sample motion from real fetal trajectories~\cite{svort}. For the clinical evaluations, we use three stacks (sagittal, coronal, and axial) per subject, automatically segment the intracranial content~\cite{monai}, and apply bias field correction~\cite{Sled1998-fn,Tustison2010-oc}, and intensity normalization as pre-processing steps. 

\begin{table}[!t]
\centering
\caption{\textbf{Performance statistics on synthetic data (N=12)}. The average and 95\% accuracy of slice alignment (TRE), and the reconstructed volume's agreement with ground truth (SSIM, NCC, PSNR) are reported. For slice pose and volume statistics, mean(std) are reported. $\uparrow$ indicates that higher values are better, $\downarrow$ indicates that lower values are better.}
\label{tab:syn_table}
\resizebox{\linewidth}{!}{
\begin{tabular}{l|cc|ccc}
\hline
& \multicolumn{2}{c|}{Slice alignment} & \multicolumn{3}{c}{Volume quality} \\
\hline
Method & TRE (mm, $\downarrow$) & TRE 95\% ($\downarrow$) & SSIM ($\uparrow$) & NCC ($\uparrow$) & PSNR ($\uparrow$) \\
\hline
SVoRT~\cite{svort} & $4.0(0.9)$ & $13.0(4.5)$ & $.87(.04)$ & $.81(.07)$ & $20.23(1.85)$ \\
NeSVoR~\cite{nesvor1} & $\mathbf{3.2}(0.7)$ & $\underline{9.2}(1.5)$ & $.87(.06)$ & $.81(.08)$ & $\mathbf{20.95}(2.17)$ \\
cSVR (ours) & $\underline{3.7}(1.7)$ & $\mathbf{6.6}(3.2)$ & $.87(.08)$ & $.81(.12)$ & $\underline{20.72}(3.42)$ \\[0.05in]
\hline
\end{tabular}
}
\end{table}

\begin{figure*}[!t]
    \centering
\includegraphics[width=\textwidth]{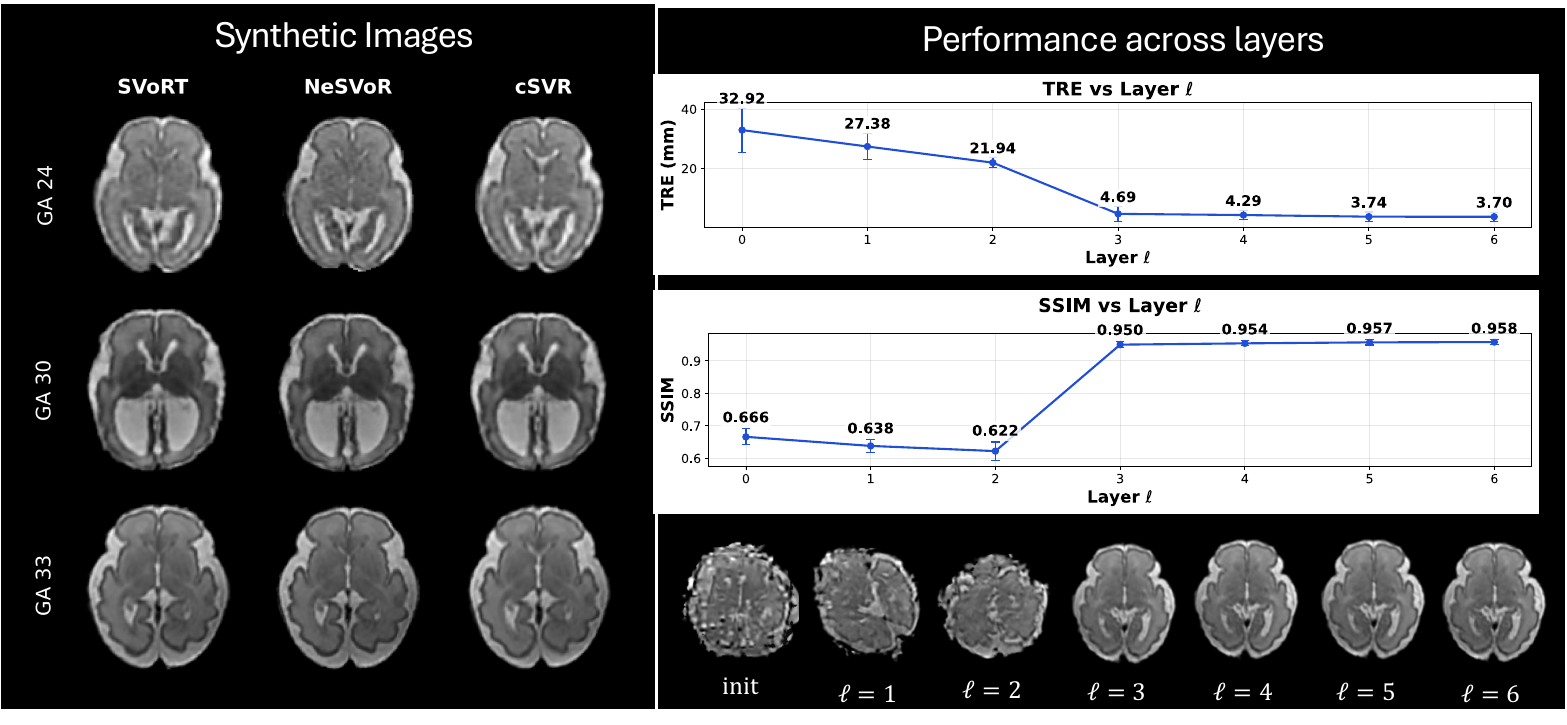}
    \caption{\textbf{Example reconstructions and TRE/SSIM analysis for different layers in the U-Net.} Left: Reconstructions on synthetic test cases show similar performance across methods. Right: Pose estimation in different layers of the network and reconstruction using estimated slice poses. The network refines the pose throughout the network, with the most significant contribution in the first three layers.}

    \label{fig:synth_brains}
\end{figure*}
\noindent
\textbf{Baseline methods.} We compare our approach with two baselines: (i) $\texttt{SVoRT}$~\cite{svort}, a transformer-based method, and (ii) $\texttt{NeSVoR}$, an implicit neural representation (INR) method. We use the latest implementation of $\texttt{SVoRT}$~\cite{svort} found in the NeSVoR repository~\cite{nesvor_repo} that employs optimization-based refinement to improve the transformer prediction. NeSVoR uses SVoRT to initialize the INR reconstruction~\cite{nesvor1}. SVoRT was trained on the FeTA dataset~\cite{feta} and shares the same synthetic evaluation set with our method. We omit classical methods, such as SVRTK~\cite{svrtk}, as NeSVoR has been shown to have comparable runtime and better reconstruction quality~\cite{masterl,nesvor1}, thus
representing the current state-of-the-art.

\noindent
\textbf{Evaluation Metrics.}
On the synthetic data, where ground truth is available, we quantify the accuracy of the pose prediction using the maximum total registration error (TRE):

\begin{equation}
\label{tre_max_in}
\operatorname{TRE}(F,F^\texttt{GT}) = \max_{p} \left\| F(p)-F^\texttt{GT}(p) \right\|_2,
\end{equation} 
\noindent where $F^\texttt{GT}$ is the ground truth slice pose. We quantify the reconstruction quality using SSIM~\cite{ssim}, PSNR, and NCC between the reconstructions and the ground truth volumes. For clinical data, we evaluate the consistency between the predicted and acquired slices using mean SSIM, PSNR, and NCC. We run all evaluations using an NVIDIA A6000 GPU.

\begin{table}[!t]
\centering
\caption{\textbf{Performance statistics for clinical data (N=18)}. The agreement between the predicted and acquired slices (SSIM, NCC, PSNR) and the runtime are reported. For image statistics, mean(std) are reported. $\uparrow$ indicates that higher values are better, 
$\downarrow$ indicates that lower values are better.
}
\label{tab:clin_table}
{\small
\begin{tabular}{l|ccc|cccc}
\hline
& \multicolumn{3}{c|}{Slice Quality} & \multicolumn{4}{c}{Runtime (seconds)} \\
\hline
Method & SSIM ($\uparrow$) & NCC ($\uparrow$) & PSNR ($\uparrow$) & Pre-processing & Reg & Recon & Total ($\downarrow$) \\
\hline
SVoRT~\cite{svort}  & $.87(.03)$ & $\mathbf{.93}(.02)$ & $23.5(1.2)$ & $11.0(2.1)$ & $\underline{2.7}(0.4)$ & $\mathbf{4.1}(0.8)$ & $\underline{17.8}(2.8)$ \\
NeSVoR~\cite{nesvor1} & $\mathbf{.91}(.03)$ & $\underline{.92}(.03)$ & $\mathbf{25.5}(2.2)$ & $11.0(2.2)$ & $\underline{2.7}(0.4)$ & $255.4(15.7)$ & $269.0(14.7)$ \\
cSVR (ours) & $\underline{.88}(.03)$ & $.91(.03)$ & $\underline{24.7}(1.2)$ & $11.1(2.2)$ & $\mathbf{1.6}(0.4)$ & $\underline{4.6}(1.1)$ & $\mathbf{17.3}(2.9)$ \\[0.05in]
\hline
\end{tabular}
}
\end{table}

\begin{figure*}[!t]
    \centering
\includegraphics[width=\textwidth]{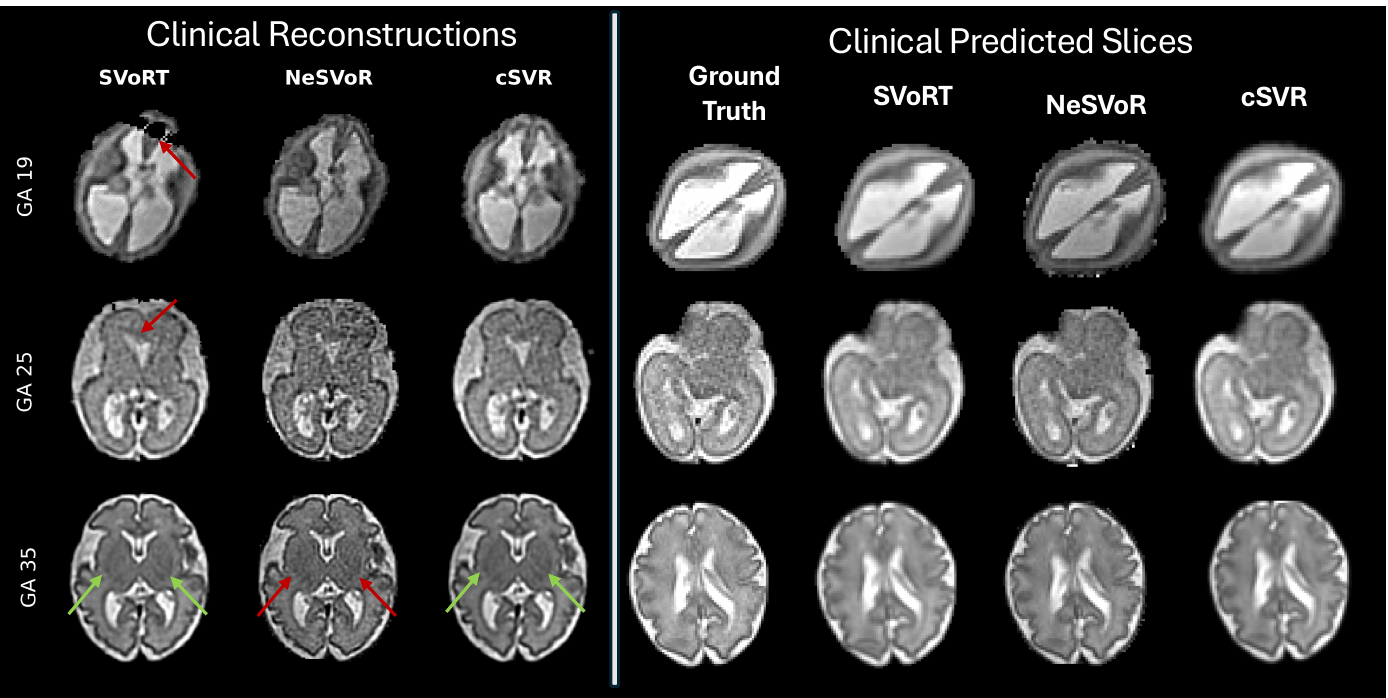}
    \caption{\textbf{Example Reconstructions for Clinical Data.} Left: Reconstructions for three real subjects. Our fast cSVR approach achieves reconstruction quality on par with state-of-the-art methods, while maintaining robustness in very early gestational age subjects while SVoRT produces holes (GA 29, red arrow). Notably, cSVR reconstruction does not include the line artifact (GA 25, red arrow) and improves delineation of cortical gray and white matter structures (GA 35, green arrows). Right: Example acquired and predicted slices for all methods. The cSVR predicted slices agree with the acquired slices and have few artifacts. }

    \label{fig:brains}
\end{figure*}

\subsection{Results}

\textbf{Synthetic Data.} Table~\ref{tab:syn_table} summarizes the statistics for synthetic data. NeSVoR achieved the best results on average slice pose prediction and volume quality, followed by our method and then SVoRT. Our method performs the best slice alignment on difficult slices (TRE 95\%), having smaller outliers for registration error. Qualitative results are visually comparable across all methods (Fig.~\ref{fig:synth_brains}, left).

We investigate the quality of the slice pose estimates for different layers of the network as seen in the right panel of Fig. \ref{fig:synth_brains} and observe that the network gradually refines the pose (TRE decreases), with large updates occurring in the early layers and finer refinements occurring gradually across multiple layers. Additionally, we report the reconstruction SSIM using pose estimates from different layers and find that the volume SSIM does not always correlate with better poses (as is the case for layers~1 and 2).

\noindent
\textbf{Clinical Data.} For clinical data, all methods achieve similar quantitative results, with NeSVoR producing the highest SSIM (Table~\ref{tab:clin_table}). Importantly, the image similarity metrics between the simulated slices and the acquired slices may not always be informative of the overall volume quality, especially in cases of poor slice coverage. This is often the case in younger subjects.  As visual inspection suggests (Fig.~\ref{fig:brains}), cSVR faithfully reconstructs a very young subject (19 weeks), while SVoRT produces holes. All methods fail to reconstruct this young subject in the canonical orientation and show some tilting, but cSVR results are the closest to the canonical pose. Additionally, cSVR shows robustness to outliers (GA 25) and good contrast to resolve gray matter with white matter (GA 35). Finally, while NeSVoR is  prone to reconstructing noisy volumes with a speckle-like pattern (GA 35), our method produces faithful reconstructions. 

Our approach substantially reduces runtime compared to state of the art (17s vs 269s) and improves registration runtime (1.6 s vs 2.7 s) compared to SVoRT \cite{svort}. This speed-up arises from fast registration enabled by the U-Net’s single-pass that incorporates multiscale unrolled optimization, in contrast to SVoRT that relies on multiple transformer iterations and additional post-processing for pose refinement.
\section{Conclusions} 
We present a fast convolutional multi-stack SVR approach that achieves significantly faster reconstruction than existing methods while maintaining comparable image quality. This work paves the way toward scanner-side clinical deployment of SVR, enhancing image interpretation and guiding acquisition in real time. The proposed network and loss function are built on generalizable field parametrization and thus can be extended to non-rigid SVR applications such as placental and whole uterus MRI. Furthermore, the pipeline generalizes to other rigid SVR scenarios involving high subject motion, including neonatal and adult patients with involuntary motion, through retraining on domain-specific data.

Future work will extend this study to a larger population, with particular emphasis on investigating the impact of mask quality and gestational age on the final reconstruction. Further work will also explore fully convolutional pose refinement via cascaded networks, with the goal of eliminating the need for classical optimization schemes.

\subsubsection*{Acknowledgments}
This work has been funded by NSF GRFP, NIH NIBIB
1R01EB036945, NIH NICHD 1R01HD114338, and NIH NIBIB 1R01EB032708.

\subsubsection{Disclosure of Interests.}
The authors have no competing interests to declare.

\newpage
\bibliographystyle{splncs04}
\bibliography{bibliography}

\end{document}